\documentclass[amsmath,amssymb,floatfix,prb,twocolumn,epsf,showpacs]{revtex4}
\usepackage{amsmath,amssymb,natbib,bm,graphicx,url,epsfig}
\usepackage[ansinew]{inputenc}

\newcommand{\be}{\begin{equation}}
\newcommand{\ee}{\end{equation}}
\newcommand{\bes}{\begin{equation}\begin{split}}
\newcommand{\ees}{\end{split}\end{equation}}

\begin{document}

\title{Planar array of semiconducting nanotubes in external electric field:
Collective screening and polarizability}

\author{T. A. Sedrakyan, E. G. Mishchenko, and M. E. Raikh}

\address{Department of Physics, University of Utah, Salt Lake
City, UT 84112}

\date{\today}

\begin{abstract}
We study theoretically the charge separation in a planar array of
semiconducting nanotubes (NTs) of  length, $2h$, placed in
external electric field, $F$,  parallel to their axes. By
employing the conformal mapping, we find  analytically the {\em
exact} distribution of the induced charge density for arbitrary
ratio of the gap, $E_g$, and the voltage drop, $eFh$, across the
NT. We use the result obtained to ({\em i}) trace the narrowing of
the neutral strip at center of the NT with increasing $F$; ({\em
ii}) analyze the polarizability of the array as a function of
$E_g$; ({\em iii}) study the field enhancement near the tips.
\end{abstract}


\maketitle

\section{Introduction}


During the past decade much progress has been achieved in growth
of vertically aligned arrays of carbon nanotubes (NTs). Vertical
alignment is attractive for such applications as field emission
displays~\cite{early0,early1,early2,early3,early4,early5,early6}
and chemical sensors\cite{sensing}. Latest
publications~\cite{latest1,latest3,latest2,latest4} report almost
perfect alignment. Since recently,  the efforts of researchers
were also directed at synthesis of nantobube arrays aligned {\em
in a plane}~
\cite{growth,growth1,ajayan00,ajayan01,ajayan02,iowa04}. The
objective of this effort is the synthesis of complex organized NT
structures for integrated molecular electronics devices and
optoelectronic applications. Growth and patterning of in-plane
oriented arrays require novel technological approaches. A number
of such approaches have been proposed in Refs.
\onlinecite{growth1,ajayan00, iowa04}, and successfully
implemented.

A planar array of NTs is shown schematically in Fig.~1. In a
typical experimental situation the distance, $d$, between the
neighboring NTs in the array is much smaller than the length,
$2h$, of the constituting NTs. This suggests that the response of
the array to the external electric field must exhibit a {\em
collective} character. For vertically oriented arrays, the
collective electrostatic response has been previously discussed in
the literature~\cite{Nilsson,wang05,we2}. This response plays a
crucial role in designing the  NT-based field emitters. For planar
arrays, being the emerging area of research, their  collective
electrostatic properties have never been addressed.

Intuitively, it is obvious that responses of vertical and planar
arrays to external electric field are dramatically different.
Indeed, external field does not penetrate into a dense enough
vertical array much deeper than the intertube distance\cite{we2}.
In other words, to the first approximation, the  tops of NTs
constituting the vertical array can be considered as a metallic
plate. By contrast, for a planar array, the force lines of
external field can extend ``above'' and ``below'' the NT plane
(see Fig.~1b), thus causing the charge separation within each NT,
constituting the array. The collective character of the
electrostatic response manifests itself in the fact that the {\em
actual field}, causing this separation in a given NT is strongly
altered by the presence of the neighboring NTs. The resulting
dipole moment, induced in an individual NT within the planar
array, depends on the charge distribution, the form of which is
not obvious {\em a priori}. The situation is even less trivial if
a planar array is made of semiconducting NTs. For a single
semiconducting NT the charge separation happens for a field
exceeding the critical value $F_{th}=E_g/2eh$.
\begin{figure}[t]
\centerline{\includegraphics[width=90mm,angle=0,clip]{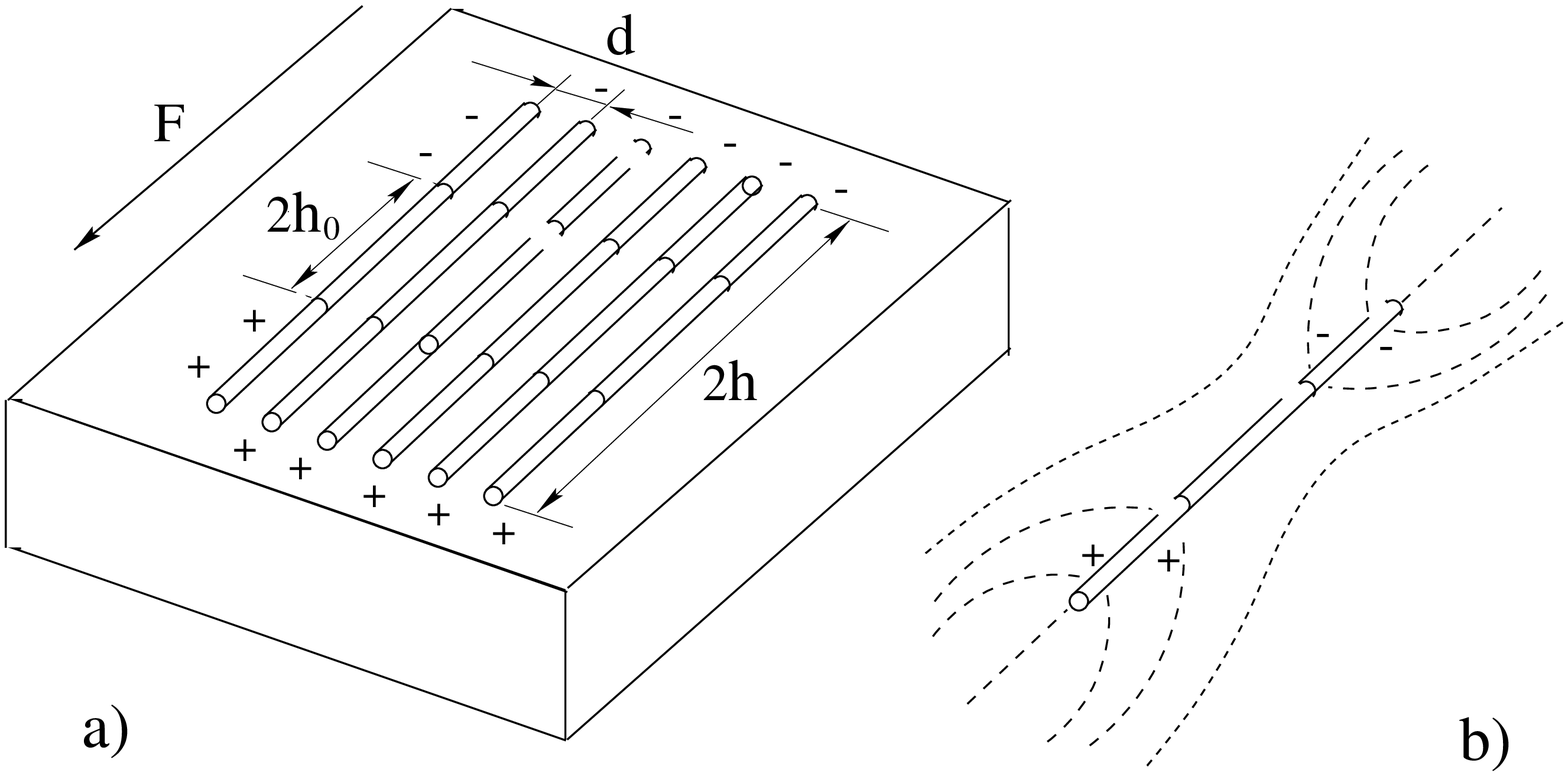}}
\caption{a) a planar array of semiconducting NTs in external
field, $F$, is shown schematically; $d$ is the distance between
the neighboring NTs.
 As a result of charge separation, the positively and negatively charged
regions of length, $h-h_0$, emerge near the tips. These regions
are separated by neutral strips of width $2h_0$; b) Force lines of
electric field above and below the array are curved as a result of
the charge separation. In charged regions the force lines enter
(exit) the NTs normally to the plane of the array.}
 \label{planar}
\end{figure}
As $F$ exceeds $F_{th}$, the charges occupy progressively larger
portion of the NT; positive and negative charges are separated by
a neutral strip~\cite{we1} of a length $E_g/eF$. It is not obvious
how, upon increasing $F$ above $F_{th}$, the field-induced charge
separation occurs in a planar array of semiconducting NTs. In
particular, how the length of the neutral strip depends on the
parameters of the array, namely, the length and density of
constituting NTs.

The questions formulated above are addressed in the present paper.
In Sect. II we extend the approach developed in
Ref.~\onlinecite{we2} to the planar array of NTs.  In Sect. III we
demonstrate that for the planar array the Thomas-Fermi equation
can be solved {\em exactly} and find a simple analytical
expression for the charge distribution along the NT for arbitrary
external field.  In particular, we find a universal relation
between the length, $2h_0$, of the neutral strip and the applied
field, $F$, and discuss qualitatively the limiting cases; Sect.
IV. In Sect. V the exact form of the charge distribution is
utilized to calculate analytically various characteristics of the
array, such as ({\em i}) electric field inside the neutral strip,
({\em ii}) electric field outside the array, ({\em iii})
enhancement of the electric field near the tips of NTs, and  ({\em
iiii})  polarizability of NTs in the array. Concluding remarks are
presented in Sect. VI.

\section{Thomas-Fermi equation for the planar array}


Derivation of the Thomas-Fermi for the planar array is completely
analogous to that for vertical array~\cite{we2} (see also
Ref.~\onlinecite{ThomasFermi}) and goes as follows.
The charge separation in a given NT of the array occurs due to the
external field and the field, created by induced charges on all
NTs. This separation proceeds until the local chemical potential
assumes a constant value along each NT of the array. This local
chemical potential of $m$-th NT, $\mu _m(z)$, is related to the
local charge density via the NT electronic spectrum\cite{we2,we1}
as follows: $ \mu _m(z)=sign(z)\large (E_g^2/4+[\rho
_m(z)/{g}]^2\large)^{1/2}$.
Then the  condition of the constant chemical potential can be
presented in the form of generalized Thomas-Fermi equation, which
for a given NT, $m$, in the array reads
\begin{eqnarray}
\label{generalized} eFz-\int_{-h}^{h}\!\!
dz^{\prime}\sum_{n}\rho_n(z^{\prime}) {\phi}(z,z^{\prime};R_m-R_n),\nonumber\\
=\text{sign}(z)~\sqrt{\frac{E_g^2}{4}+\Biggl[\frac{
\rho_m(z)}{g}\Biggr]^2},
\end{eqnarray}
where $|R_n-R_m|$ is the distance between the centers of the NTs
$n$ and $m$. In Eq.~(\ref{generalized}) the
dimensionless interaction parameter $g$ is related to the electron
velocity, $v_0$, in the NT as\cite{we1,we2} $g=({2Ne^2}/{\pi
\epsilon^{\ast}\hbar v_0})$, where $N$ is the number of conducting
channels, which assumes the values $2$ or $4$ depending on the
spin and band degeneracy. This degeneracy is governed by the NT
chirality. The potential, ${\phi}(z,z^{\prime})$, in
Eq.~(\ref{generalized}) is given by
\begin{eqnarray}
{\phi}(z,z^{\prime};R)&=&\frac{e}{\epsilon^{\ast}\sqrt{(z-z^{\prime})^2+
R^2}},
\end{eqnarray}
where $\epsilon^{\ast}=(\epsilon +1)/2$ is the average dielectric
constant between the substrate and the air.
For a regular array we set all $\rho_n(z)$ equal to $\rho(z)$ and
introduce the NT density, ${\cal N}_0$, as ${\cal N}_0=1/d$. As
the next step, we take the continuous limit of
Eq.~(\ref{generalized}). In doing so, we first isolate the NT with
$n=0$ and rewrite Eq.~(\ref{generalized}) in the form
\begin{eqnarray}
\label{discrete}
eFz&=&\text{sign}(z)~\sqrt{\frac{E_g^2}{4}+\Biggl[\frac{
\rho(z)}{g}\Biggr]^2}+2{\cal L}_d\rho(z)\nonumber\\
&+&\frac{{\cal
N}_0}{\epsilon^{\ast}}\int_{0}^h\!dz^{\prime}\rho(z^{\prime}){\cal
S}(z,z^{\prime}),
\end{eqnarray}
where $R_n=nd$, and the kernel, ${\cal S}(z,z^{\prime})$, is
defined as
\begin{eqnarray}
\label{kernel} {\cal S}(z,z^{\prime})=\sum_{n\neq 0}\Biggl[
\frac{1}{\sqrt{(z-z^{\prime})^2+R_n^2}}-
\frac{1}{\sqrt{(z+z^{\prime})^2+R_n^2}}\Biggr].\nonumber\\
\end{eqnarray}
Here we have used the fact that $\rho(z)=-\rho(-z)$. The term
$2{\cal L}_d\rho(z)$ in the rhs of Eq.~(\ref{discrete}) comes from
the term with $n=0$ in the sum (\ref{generalized}) and describes
the ``self-action'' of a given $NT$. For an isolated NT\cite{we1}
this term has the form $2\ln(h/r)\rho(z)$, where $r$ is the NT
radius. In the presence of neighboring NTs, the self-action is
limited by the distance $\sim d$, so that ${\cal L}_d$ is equal to
$\ln(d/r)$ analogously to the vertical array\cite{we2}. Now taking
the continuous limit in Eq.~(\ref{discrete}) reduces to
replacement the sum over $n$ in the kernel Eq.~(\ref{kernel}) by
the integral, leading to the following integral equation
\begin{eqnarray}
\label{inteq1}
eFz&=&\text{sign}(z)~\sqrt{\frac{E_g^2}{4}+\Biggl[\frac{
\rho(z)}{g}\Biggr]^2}\nonumber\\ &+& 2{\cal L}_d\rho(z)+ \frac{ 2
{\cal
N}_0}{\epsilon^{\ast}}\int_{-h}^h\!dz^{\prime}\rho(z^{\prime})\ln
(\vert z-z^{\prime}\vert).
\end{eqnarray}
Rigorous justification of the continuous description is given in
Appendix, where we show that the effect of discreteness of NTs in
the array amounts to a small $\sim d/h$ correction to $\rho(z)$.

\section{Solution of the integral equation}
\subsection{Metallic NTs}
In the case of metallic NTs we set $E_g=0$ in Eq.~(\ref{inteq1}).
We will also neglect the self-action term, $2{\cal L}_d$, in the
rhs; the validity of this step will be verified in the end of this
Subsection.  Then, for positive $z$, Eq.~(\ref{inteq1}) takes the
form
\begin{eqnarray}
\label{bulkeq1} eFz=\frac{2 {\cal
N}_0}{\epsilon^{\ast}}\int_{0}^h\!dz^{\prime}\rho(z^{\prime})\ln\Bigg|\frac{z+z^{\prime}}{z-z^{\prime}}\Bigg|.
\end{eqnarray}
Note that  Eq.~(\ref{bulkeq1}) does not contain any small
parameter. Indeed, upon introducing dimensionless charge density
and coordinate as
\begin{equation}
\label{dimensioless} \tilde\rho=\frac{{\cal
N}_0}{e\epsilon^{\ast}F}\;\rho,~~~~~~{\tilde z}=\frac{z}{h},
\end{equation}
Eq.~(\ref{dimensioless}) assumes the following form
\begin{equation}
\label{dimensionless1} {\tilde z}=\int_0^1\!d{\tilde
z}^{\prime}\;\tilde\rho({\tilde z}^{\prime})\;
\ln\Bigg|\frac{{\tilde z}+{\tilde z}^{\prime}}{{\tilde z}-{\tilde
z}^{\prime}}\Bigg|.
\end{equation}
In Eq.~(\ref{dimensionless1}) the relevant values of $\tilde\rho$,
$z^{\prime}$, and ${\tilde z}$ are all of the order of unity. It
turns out, however, that, despite the absence of a small
parameter, Eq.~(\ref{dimensionless1}) can be solved analytically
in the closed form. The solution reads
\begin{equation}
\label{dimensionless2} {\tilde \rho}(\tilde
z^{\prime})=\frac{{\tilde z}^{\prime}}{\pi\sqrt{1-({\tilde
z}^\prime)^2}}.
\end{equation}
In order to check that Eq.~(\ref{dimensionless2}) is indeed the
solution of the integral equation (\ref{dimensionless1}), we
substitute (\ref{dimensionless2})    into the rhs of
Eq.~(\ref{dimensionless1})
and perform integration by parts as follows
\begin{eqnarray}
\label{dimensionless3} \int_0^1\!d{\tilde z}^{\prime}\frac{{\tilde
z}^{\prime}}{\sqrt{1-({{\tilde z}^{\prime}})^2}}
\ln\Bigg|\frac{{\tilde z}+{\tilde z}^{\prime}}{{\tilde z}-{\tilde z}^{\prime}}\Bigg|\qquad\qquad\qquad\\
=-2{\tilde z}\int_0^1\!d{\tilde z}^{\prime}\frac{\sqrt{1-({{\tilde
z}^{\prime}})^2}}{({\tilde z}^{\prime})^2 -{\tilde z}^2}
=-2{\tilde
z}\int_0^{\pi/2}\!d\varphi\frac{(\cos\varphi)^2}{(\sin\varphi)^2-{\tilde
z}^2},\nonumber
\end{eqnarray}
where the integration in the rhs implies taking the principal
value.
 The last identity in Eq.~(\ref{dimensionless3}) emerges upon  substitution ${\tilde z} =\sin\varphi$.
Our key observation is that the integral in rhs is equal $-\pi/2$
for {\em all} $\vert {\tilde z}\vert < 1$, so that
$\tilde\rho({\tilde z})$ in the form (\ref {dimensionless2}) is
indeed the solution of Eq.~(\ref{dimensionless1}).
Returning to dimensional variables, we present the  final result
for the charge density distribution in the form
\begin{eqnarray}
\label{metalsol} \rho(z)=\frac{e\epsilon^{\ast}Fz}{\pi{\cal
N}_0\sqrt{h^2-z^2}}.
\end{eqnarray}
Recall now, that this result was obtained neglecting the
self-action. To verify the validity of this assumption, we
estimate the ratio, $2{\cal L}_d\rho(z)/e\epsilon^{\ast}Fz$, of
the self-action term in Eq.~(\ref{inteq1}) to the lhs. As follows
from Eq.~(\ref{metalsol}), this ratio is $\sim {\cal L}_d/{\cal
N}_0h$. Thus, the self-action can be neglected,  if ${\cal N}_0 >
{\cal L}_d/h$, i.e., when the array is sufficiently dense.
Although the self-action term is relatively small, it overweighs
the integral term near the tips, where
the solution Eq.~(\ref{metalsol}) diverges.
This suggests that the divergence in (\ref{metalsol}) must be cut
off at $(h-z) \sim h\left({\cal L}_d/h{\cal N}_0\right)^2$, where
the the first term in the rhs of Eq.~(\ref{inteq1}) becomes
comparable to the lhs. At the cutoff, the ratio $(h-z)/h$ is $
\sim \left({\cal L}_d/h{\cal N}_0\right)^2 \ll 1$, indicating
that, in a dense array, the solution Eq.~(\ref{metalsol}) applies
up to the close vicinity of the tip.


\subsection{Semiconducting NTs}

\subsubsection{\bf Reduction of the Thomas-Fermi equation to the Hilbert form}
Upon neglecting self-action, and in the limit of large $g$, the
Thomas-Fermi equation for the charge distribution takes the form
\begin{eqnarray}
\label{bulkeq} eFz-\text{sign}(z)~\frac{E_g}{2}=\frac{2 {\cal
N}_0}{\epsilon^{\ast}}\int_{-h}^h\!dz^{\prime}\rho(z^{\prime})
\ln(\vert z-z^{\prime}\vert).
\end{eqnarray}
As was explained qualitatively in the Introduction [see also Ref.
(\onlinecite{we1})], in contrast to the metallic case, the
solution of Eq.~(\ref{bulkeq}) is singular. This is because the
positive and negative charges near the tips, induced by the
external field, are separated by the neutral strip, caused by a
finite bandgap, $E_g$, with boundaries at $z=\pm h_0$. Within this
strip we have $\rho(z)=0$. As the external field increases, the
strip narrows. Solving Eq.~(\ref{bulkeq}) implies finding both the
form of $\rho(z)$ outside the interval $[-h_0, h_0]$ and the
dependence of $h_0$ on the external field.

Formally, with finite $E_g$, the integral equation contains one
dimensionless parameter, $E_g/eFh$. In this subsection we
demonstrate that the exact solution of Eq.~(\ref{bulkeq})  can be
obtained for {\em arbitrary} value of this parameter. As a first
step it is convenient to differentiate both sides of
Eq.~(\ref{bulkeq}) with respect to $z$. This yields the following
Hilbert-type integral equation
\begin{eqnarray}
\label{hilbert} \frac{e\epsilon^{\ast}F}{2 {\cal
N}_0}=\int_{-h}^{-h_0}\!dz^{\prime}\frac{\rho(z^{\prime})}{z-z^{\prime}}+\int_{h_0}^{h}
\!dz^{\prime}\frac{\rho(z^{\prime})}{z-z^{\prime}}.
\end{eqnarray}\\
Our key observation is that the 2D geometry of the array allows
one to employ the technique of conformal mapping in order to solve
Eq.~(\ref{hilbert}). The choice of the appropriate conformal
transformation is described below.

\subsubsection{\bf Conformal Mapping}

 The conformal mapping method can be applied
for large $g$, when
each of the regions $-h <z< -h_0$ and $h_0 <z <h$
are {\em equipotential}. Then a general solution of the 2D Poisson
equation with boundary conditions of equipotentiality on flat
metal surfaces can be obtained from mapping of the  boundaries of
these surfaces on the real axis in the complex plane. What makes
the case of semiconducting array nontrivial, is that the
boundaries of equipotential regions have a doubly-connected
geometry.

To gain an insight as to what is the appropriate form of the
transformation for the
geometry under study, it is instructive to start with an auxiliary
problem of two {\em isolated}  metallic plates in external field,
occupying the intervals $[-h,-h_0]$ and $[h_0,h]$. Two metallic
plates correspond to the two cuts on the complex plane,
as shown in Fig.~2. The general solution of Eq.~(\ref{hilbert})
for this geometry can be found in the textbooks
\cite{muskhelishvili,nehari, schwarz-chrisoffel,conf-map}. It can
be expressed through the integrals along the closed contours
$L_{-}$ and $L_{+}$, encompassing the cuts in the clockwise
direction, in the following way
\begin{eqnarray}
\label{cont} \qquad\rho(z)=\frac{A}{\sqrt{(z^2-h_0^2)(h^2-z^2)}}\qquad\qquad\qquad\qquad\qquad\nonumber\\
\times\Biggl\{\oint_{L_-\cup
L_{+}}\!\!dt\;\frac{\sqrt{h^2-t^2}\;\sqrt{t^2-h_0^2}}{z-t}\Biggr\}.\qquad
\end{eqnarray}

\begin{figure}[t]
\centerline{\includegraphics[width=90mm,angle=0,clip]{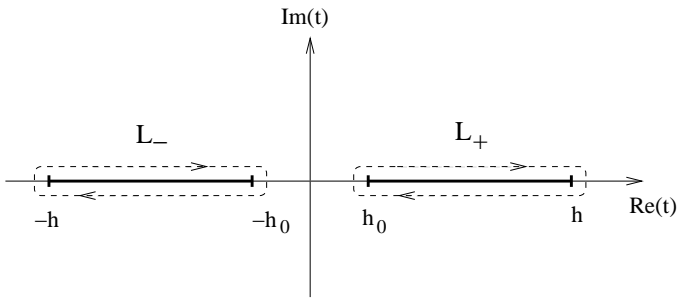}}
\caption{Contours of integration in Eq.~(\ref{cont}), $L_{-}$ and
$L_{+}$, encompassing the cuts in the clockwise direction. }
\label{fig:1}
\end{figure}


\subsubsection{\bf Solution for the charge density distribution}

Although Eq.~(\ref{cont}) satisfies Eq.~(\ref{hilbert}), it cannot
be applied to the planar array of semiconducting NTs. The reason
lies in the  fundamental difference between the isolated metallic
plates and array of semiconducting NTs. In the former case {\em
each} plate respects  electroneutrality, i.e.,
$\int_{-h}^{-h_0}dz\rho(z)=0$, ~ $\int_{h_0}^{h}dz\rho(z)=0$,
while the array is neutral only {\em as a whole}. Still
Eq.~(\ref{cont}) can be modified to account for this difference.
The modification amounts to adding a constant to the integral in
Eq.~(\ref{cont}) after which acquires the following form
\begin{eqnarray}
\label{chargedens} \rho(z)=\frac{1}{\sqrt{(z^2-h_0^2)(h^2-z^2)}}\qquad\qquad\qquad\qquad\qquad\nonumber\\
\times \text{sign}(z)\Biggl\{2A\int_{h_0}^h\!dt\;
\frac{t\sqrt{h^2-t^2}\;\sqrt{t^2-h_0^2}}{t^2-z^2}+C\Biggr\}.\quad
\end{eqnarray}
It can be checked by a direct substitution that
Eq.~(\ref{chargedens}) is the solution of Eq.~(\ref{hilbert}) for
arbitrary $C$. On the other hand, with nonzero $C$, the net charge
in each of the intervals $[-h,-h_0]$ and $[h_0,h]$ is finite. On
the physical grounds, the value of the constant $C$ should be now
determined from  the condition $\rho(h_0)=0$, which fixes the
ratio of the constants $C$ and $A$ at the value $C/A
=-(\pi/2)(h^2-h_0^2)$. The meaning of this condition is that the
charge density vanishes at the boundaries of the neutral strip. It
turns out that the result Eq.~(\ref{chargedens}) can be greatly
simplified. Namely, the integral can be evaluated {\em
analytically} leading to the following simple expression for the
induced charge density
\begin{eqnarray}
\label{semicond}
\rho(z)=\text{sign}(z)~\frac{e\epsilon^{\ast}F}{\pi{\cal
N}_0}\sqrt{\frac{z^2-h_0^2}{{h^2-z^2}}}.
\end{eqnarray}
 The relation between $h_0$ and $E_g$
can be now established by substitution of the solution
Eq.~(\ref{chargedens}) with unknown constants $A$ and $h_0$ into
Eq.~(\ref{hilbert}) and equating linear in $z$ and constant terms
to $eFz$ and $E_g/2$, respectively. This yields
$A=-\epsilon^{\ast}F/\pi^2{\cal N}_0$ and the following
transcendental equation for $h_0$
\begin{eqnarray}
\label{gap} \frac{E_g}{2eFh}={\bf \text{\large
E}}\left[\frac{h_0^2}{h^2}\right]-
\left(1-\frac{h_0^2}{h^2}\right){\bf \text{\large
K}}\left[\frac{h_0^2}{h^2}\right],
\end{eqnarray}
where  ${\bf \text{\large K(x)}}$  and  ${\bf \text{\large E(x)}}$
are the elliptic integrals of the first and the second kind,
respectively.
%
%
\begin{figure}[th]
\centerline{\includegraphics[width=90mm,angle=0,clip]{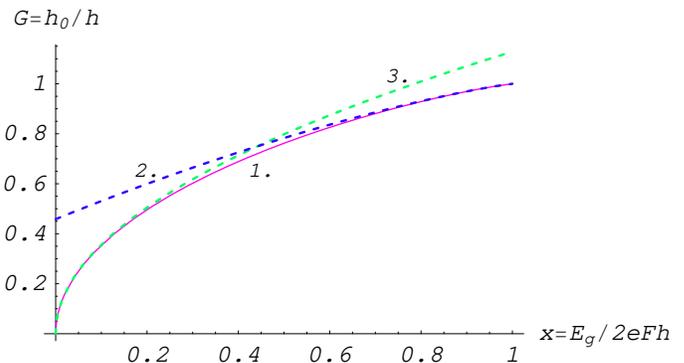}}
\caption{(Color online) Curve 1: Dimensionless length of a neutral
strip $G(x)=h_0/h$ is plotted versus dimensionless ratio
$x=F_{th}/F$ from Eq.~(\ref{gap}); Curves 2 and 3 are the
asymtotes plotted from Eq.~(\ref{G_asympt}).  } \label{universal}
\end{figure}
%
In conclusion of this Section, we note that conformal mapping was
previously employed to find the positions of boundaries of the
two-dimensional electron gas in a smooth external potential
\cite{glazman91} as well as to find the positions of edges of
incompressible strips in the quantum Hall sample
\cite{chklovskii}.

\section{Discussion}

Eq.~(\ref{gap}) defines a universal dependence of the length,
$2h_0$, of the neutral strip on the applied field, $F$. This
dependence can be parameterized as
\begin{eqnarray}
\label{G} h_0=h\;G\Bigl(\frac{F_{th}}{F}\Bigr),
\end{eqnarray}
where $F_{th}= E_g/2eh$ is the threshold value of $F$ that causes
the charge separation in the array. The dimensionless function,
$G(x)$  is shown in Fig.~2. It has the following asymptotes
\begin{eqnarray}
\label{G_asympt}
G(x)&=&\frac{2\sqrt{x}}{\sqrt{\pi}}, \;\;\;\;\;\; x\ll 1; \nonumber\\
G(x)&=&1+\frac{2\;(1-x)}{\ln\;(1-x)}, \;\;\;(1-x)\ll 1.
\end{eqnarray}
As seen from Fig.~2, the above asymptotes approximate $G(x)$ very
closely. The dependence $G(x)$ should be contrasted to the
corresponding dependence for an isolated NT, where it has the
form\cite{we1} $G_0(x)=x$.  We note,  that $G(x)>G_0(x)$ for all
$x$. In other words, the neutral strip for NT array is {\em wider}
than that for a single NT. The difference is especially pronounced
for strong fields, corresponding to $x\ll 1$, where we have
$G(x)/G_0(x) \sim x^{-1/2}$. The origin of such a difference lies
in the fact that the array screens the external field much more
efficiently than a single NT. As a result, for strong fields, when
$h_0\ll h$, the field inside the strip is $\sim (h_0/h)$ times
smaller than $F$. Then the condition that the potential drop
within the neutral strip is equal to $E_g$ can be presented as
$eh_0(h_0F/h)\sim E_g$. This condition yields $h_0\sim
\bigl(E_gh/eF\bigr)^{1/2}$, which is in agreement with the
small-$x$ asymptote $G(x)\sim x^{1/2}$ of Eq.~(\ref{G_asympt}).


For an individual NT, the induced charge density changes linearly
near the edges of the neutral strip \cite{we1}. By contrast, for a
2D array, $\rho(z)$ described by Eq.~(\ref{semicond}) exhibits a
singular behavior, $\rho(z) \propto \pm \sqrt{z \pm h_0}$, near
the boundaries of the neutral strip.

The principal assumption that allowed us to find the charge
distribution in semiconducting NT analytically is that in
Eq.~(\ref{bulkeq}) we had replaced $\sqrt{(E_g/2)^2+(\rho /g)^2}$
in Eq.~(\ref{inteq1}) by $E_g/2$. Using the result
Eq.~(\ref{semicond}), we can now check the validity of this
assumption. It is easy to see that the replacement is justified
when the external field is not too strong, so that the
dimensionless ratio  $x=F_{th}/F$ in Eq.~(\ref{G_asympt}) must
exceed the value $(g{\cal N}_0h)^{-1}\ll 1$.

\section{Implications}

Exact results Eqs.~(\ref{metalsol}), (\ref{semicond}) for the
charge distribution in metallic and semiconducting arrays,
obtained above, enable us to calculate analytically various
observable characteristics of the array. In this Section we will
focus on the following  characteristics: electric field
distribution inside and outside of the array, and collective
polarizability of the array. Concerning the electric field
distribution, the questions most interesting for applications are
the distribution of electric field within the neutral strips and
behavior of electric field near  the NT tips. The fist question is
important for electrooptical measurements, while the second
question is crucial for the field emission. In addition, it is
instructive to study quantitatively how the charges, induced in
the NTs by external field, alter this field in the regions above
and below the array. In particular, the interesting question is
how the spatial decay of the electric field component normal to
the array depends on the position, $z$, along the NTs. These
questions are addressed in subsections that follow.

\subsection{Electric field inside the neutral strip}

Electrostatic potential inside the neutral strip is given by
superposition of the external potential, $eFz$, and the induced
potential
\begin{eqnarray}
\label{pot_inside} \phi^{ind}(z)=\frac{ {2\cal
N}_0}{e\epsilon^{\ast}}\int_{h_0}^h\!\!dz^{\prime}\rho(z^{\prime})
\ln\Bigg|\frac{z+z^{\prime}}{z-z^{\prime}}\Bigg|,
\end{eqnarray}
where $z<h_0$. Upon substituting   the exact form of the charge
density, Eq.~(\ref{semicond}), in to Eq.~(\ref{pot_inside}) the
integration can be performed analytically. We will present the
result for the tangential component of the net electric field
which is the sum of the external, $F$, and induced,
$F_z^{ind}=\partial\phi^{ind}(z)/\partial z$, fields inside the
neutral strip,
\begin{eqnarray}
\label{f_inside}
F_z^{st}=F_z^{ind}+F=F\sqrt{\frac{\bigl[h_0(F)\bigr]^2-z^2}{h^2-z^2}},
\end{eqnarray}
where the dependence $h_0(F)$ is defined by Eq.~(\ref{gap}) and
plotted in Fig.~3. In the limit of weak external field, close to
the threshold  value, $F_{th}$, Eq.~(\ref{f_inside}) yields a
natural result that $F_z^{st}\approx F$. This is because the
induced charge occupies only a small portion, $(h - h_0)\ll h_0$,
of the NTs. In the opposite limit of the strong external field,
$F\gg F_{th}$, the field, $F_z^{st}$, at the center of the NTs is
equal to $Fh_0[F]/h \ll F$, i.e., it is strongly suppressed, and
falls off towards the ends of the strip. Note, that in our
qualitative derivation of the width of the neutral strip we have
used the estimate $F_z^{st}\sim  Fh_0/h$ for the field inside the
strip. Therefore, Eq.~(\ref{f_inside}) justifies this estimate.
The reason why we focused on the tangential component, $F_z^{st}$,
is its possible  relevance for the experiment\cite{kennedy05} on
electroabsorption of light. It might seem that, with photon energy
$\sim 1$eV much bigger than $eFh$, the large-scale nanotube
geometry is not important.  This, however, is not the case. The
reason is that the dipole moment of the many-body optical
transition
 is directed {\em along} the tube \cite{Kane,Spataru,Perebeinos,Liang,exciton}.
In analysis of the data in Ref. \onlinecite{kennedy05} it was
assumed that in the change,
$[\alpha(\omega,F_z)-\alpha(\omega,0)]$, of the absorption
coefficient, $F_z$ is the {\em external} field. Meanwhile, the
optical transition is modified by the {\em acting} field,
$F_z^{st}$. Then Eq.~(\ref{f_inside}) suggests that the acting
field can be much smaller than the external field and, moreover,
behaves as $F_z^{st}\propto F^{1/2}$ due to the  collective
screening.

\subsection{Electric field outside the array}

Consider a point with coordinate, $z$, located at a  distance,
$H$, above the array.  Similarly to Eq.~(\ref{f_inside}), the
parallel and perpendicular components of the acting field can be
evaluated from
\begin{eqnarray}
\label{components} F_z(z,H)\!\!&=&\!\!F-\frac{\partial}{\partial
z}\int_{-h}^{h}\!\!dz^{\prime}\!\!\int_{-\infty}^{\infty}\!\!dx\frac{
{\cal N}_0\rho(z^{\prime})}
{\Bigl[(z^{\prime}-z)^2 +x^2+H^2\Bigr]^{1/2}},\nonumber\\
F_{\perp}(z,H)\!\!&=&\!\!-\frac{\partial}{\partial
H}\int_{-h}^{h}\!\!dz^{\prime}\!\!\int_{-\infty}^{\infty}\!\!dx\frac{
{\cal N}_0\rho(z^{\prime})} {\Bigl[(z^{\prime}-z)^2
+x^2+H^2\Bigr]^{1/2}}.\quad\nonumber\\
\end{eqnarray}
We will restrict ourselves to metallic NTs. In this case the
integration in Eq.~(\ref{components}) can be performed
analytically, yielding
\begin{eqnarray}
\label{fields}
F_z=\frac{F}{\sqrt{2\Lambda}}\sqrt{(z^2+H^2)\sqrt{\Lambda}+\Lambda -h^2(h^2+H^2-z^2)},\nonumber\\
F_{\perp}=\frac{\sqrt{2}F}{\sqrt{\Lambda}}\sqrt{(z^2+H^2)\sqrt{\Lambda}-\Lambda+h^2(h^2+H^2-z^2)},
\nonumber\\
\end{eqnarray}
where we had introduced an auxiliary function $\Lambda (z,H)$
defined as follows
\begin{eqnarray}
\Lambda (z,H) =(H^2+z^2-h^2)^2+4h^2H^2.
\end{eqnarray}
Eq.~(\ref{fields}) allows one to trace how the tangent component,
$F_z$, grows from zero to $F$ with increasing $H$, while the
normal component falls off from $2\pi{\cal N}_0\rho(z)$ to zero
with increasing $H$. The characteristic scale of change for both
components is $H \sim h$. This behavior is illustrated in Fig.~4.
\begin{figure}[th]
\centerline{\includegraphics[width=90mm,angle=0,clip]{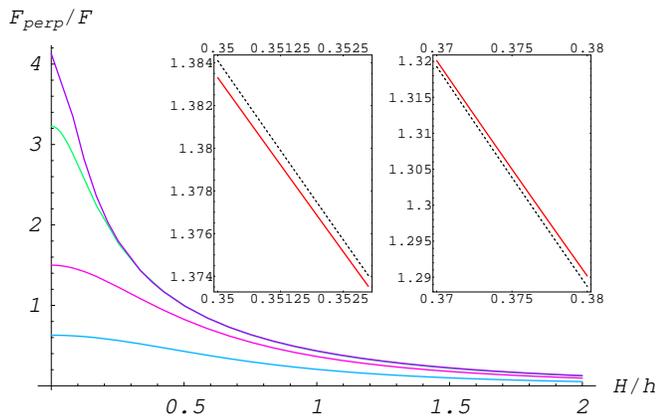}}
\caption{ (Color online) Dimensionless normal component,
$F_{\perp}/F$, of the net electric field is plotted from
Eq.~(\ref{fields})
vs. dimensionless distance, $H/h$, from the array for various
values of coordinate, $z$, along the array. The plots are
presented from bottom up: $z/h=0.3;\;0.6;\;0.85;\;0.9.$ Two insets
illustrate that the curves $F_{\perp}(H)$ for $z/h =0.8$ and
$z/h=0.9$ intersect each other. The actual point of intersection
(not showed)
 is at $H/h \approx 0.3677$.}\label{Fp}
\end{figure}
\begin{figure}[th]
\centerline{\includegraphics[width=90mm,angle=0,clip]{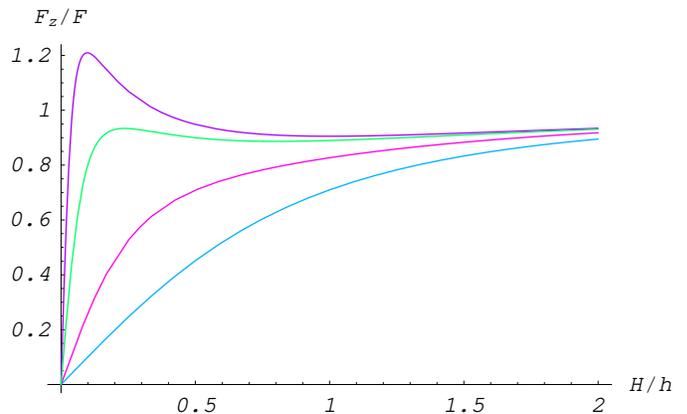}}
\caption{(Color online) Dimensionless tangent component,
$F_{z}/F$, of the net electric field is plotted from
Eq.~(\ref{fields})
vs. dimensionless distance, $H/h$, from the array for various
values of coordinate, $z$, along the array. The plots are
presented from bottom up: $z/h=0.1;\;0.7;\;0.9;\;0.95.$}
\label{Fz}
\end{figure}

We see that, as could be expected, $F_{\perp}$ falls off
monotonically with increasing the distance, $H$, from the array.
Less trivial is that the curves for $z/h =0.85$ and $z/h =0.9$
intersect at $H/h \approx 0.3677$ (see inset in Fig.~5). This
implies that the derivative $\partial F_{\perp}/\partial z$ is
zero at this point. On the other hand the relation  $\partial
F_{\perp}/\partial z =\partial F_{z}/\partial H $ suggests that
$F_z(H)$ must have a maximum at $H/h \approx 0.3$. As seen from
Fig.~5, this is indeed the case. It is also seen from Fig.~4 that,
for $z/h$ approaching  $1$, the field $F_{\perp}$ grows in the
vicinity of the plane of the array $H \rightarrow 0$. This is the
manifestation of the enhancement of electric field near the tip.
We address this issue in more detail in the next subsection.

\subsection{Field enhancement near the tips}

Consider a point, $z_0$, on the axis of one of the NTs outside the
array, $z_0>h$. The field in this point has only $z$-component
given by
\begin{eqnarray}
\label{pot} F_z^{ind}=\frac{ {\cal
N}_0}{e\epsilon^{\ast}}\int_{0}^h\!dz\rho
(z)\Biggl(\frac{1}{z_0+z}-\frac{1}{z_0-z}\Biggr).
\end{eqnarray}
With metallic charge density distribution Eq.~(\ref{metalsol}) the
integration in Eq.~(\ref{pot}) can be performed analytically. We
will present the result for the field enhancement factor, $\beta =
F_z^{ind}/F$, which is the conventional characteristics of the
field near the tip
\begin{eqnarray}
\label{F_induced} \beta = \frac{
h^2}{\sqrt{z_0^2-h^2}\;(z_0+\sqrt{z_0^2-h^2})}.
\end{eqnarray}
We can now compare the dependence $\beta(z_0)$ for the  planar
array and for an isolated NT.
 For the  isolated NT we have \cite{we1}
$\beta(z_0) \propto h/(z_0-h)$ (in the present notations). By
contrast, Eq.~(\ref{F_induced}) yields a slower decay,
$\beta(z_0)\sim h^{1/2}/(z_0-h)^{1/2}$. On the other hand, the
magnitude of the enhancement factor is much weaker for the planar
array than for the isolated NT. Obviously, this reduction is due
to the collective screening.

\subsection{Induced dipole moment}
The dipole moment, $P$, of a given NT within the array is defined
as
\begin{eqnarray}
\label{dipole} P(F)=2\int_{h_0}^h\!\!dz\;z\rho(z).
\end{eqnarray}
Analytical evaluation of the integral in Eq.~(\ref{dipole}), with
the charge density (\ref{semicond}), yields the following
expression for the dipole moment
\begin{eqnarray}
\label{DD} {P}(F)=\frac{e\epsilon^{\ast}F}{4{\cal
N}_0}\bigl(h^2-\left\{h_0(F)\right\}^2\bigr).
\end{eqnarray}
It is instructive to consider the limiting cases of weak and
strong fields. Using the low-field and strong-field asymptotes of
$h_0(F)$ given by Eq.~(\ref{G_asympt}), we obtain
\begin{eqnarray}
\label{dip_1}
{P}(F)\!\!&=&\!\!\Biggl(\frac{e\epsilon^{\ast}F_{th}h^2}{{\cal
N}_0}\Biggr)\frac{1-\frac{F_{th}}{F}}{\vert
\ln\left[1-\frac{F_{th}}{F}\right]\vert},
\;\;\;(F-F_{th})\ll F_{th},\nonumber\\
{P}(F)\!\!&=&\!\!\frac{e\epsilon^{\ast}Fh^2}{4{\cal N}_0}\Biggl(1-
\frac{4F_{th}}{\pi F}\Biggr), \qquad F \gg F_{th}.
\end{eqnarray}
Note that in the limit of low fields, $(F-F_{th})\ll F_{th}$, the
behavior of $P(F)$ is singular, unlike the individual NT, where
one has\cite{we1} $P(F)\propto \left[1-F/F_{th}\right]^2$. This
singularity is the manifestation of the singular behavior of the
charge density, $\rho(z)\propto \sqrt{z-h_0}$,
 at the boundary of the neutral strip. By contrast, for the
individual NT, this behavior is regular, namely $\rho(z) \propto
(z-h_0)$.

In the strong-field limit, the contribution of the neutral strip
to $P(F)$ constitutes a relatively small correction $\sim
F_{th}/F$. Essentially, the polarizability $\chi = P(F)/F$ of a
given NT is $\sim eh^2/{\cal N}_0$, as seen from Eq.~(\ref{DD}).
On the other hand, the value of $\chi$ for  the individual NT can
be estimated from the textbook problem\cite{Landau84} of
polarization of an ellipsoid with large ratio of axes in electric
field directed along the major axis. With   logarithmic accuracy
one has, $\chi \sim eh^3$. Thus, compared to the individual NT,
the polarizability of a given NT within the array is suppressed by
a large factor ${\cal N}_0h$. This conclusion can be reformulated
as follows. An element of the array of a width $h$ contains ${\cal
N}_0h$ NTs. The integral polarizability of this element is $\sim
({\cal N}_0h)(eh^2/{\cal N}_0)=eh^3$, i.e., it is of the same
order as polarizability of the individual NT. Therefore, the
magnitude of suppression of $\chi$ within the array, can be
understood from the  fact that, within the array, the constituting
NTs ``talk to each other'' only if their separation is smaller
than $h$.


\section{Concluding remarks}

Throughout the paper we assumed that the NTs in the planar array
are parallel to each other. Note, that recently a different class
of planar NT arrays in the form of a {\em random network},
residing on the polymeric substrate, and connecting the source and
drain electrodes has been discussed and realized
experimentally~\cite{network1,network2,network3}. It was
demonstrated that such a network can emulate the channel of the
field-effect transistor. Since in our consideration of the planar
array was carried in continuous limit, the results obtained are
relevant to the random planar dense arrays.

Our main results are Eqs.~(\ref{metalsol}), (\ref{semicond})
describing the distribution of induced charge in metallic and
semiconducting planar arrays, respectively. These results were
obtained in the continuous limit, i.e., we replaced the array of
individual NTs with a two-dimensional continuous layer of a zero
width. Such a replacement is justified with high accuracy as long
as the NT length, $h$, exceeds the separation, $d$, between the
neighboring NTs. More quantitatively, as shown in Appendix, the
relative correction to the induced charge distribution is $\sim
d/h$. Similarly, the randomness in distances between the
neighboring NTs is not important for the planar array. This should
be contrasted to the {\em infinite} vertical array\cite{we2},
where the distance between {\em neighboring} NTs governs the
penetration depth of the external field. By contrast, the
continuous description adopted in the present paper applies with
high accuracy, since, in the planar array, the induced charge
extends over  the entire length, $2h$, of the NTs. This length
the is much bigger than ${\cal N}_0^{-1}$. The discreteness of the
array manifests itself only in the close vicinity of the tips,
where it cuts off the singularity $\rho(z) \propto (h-z)^{-1/2}$
in the distribution of the induced charge.

It is instructive to trace how the results obtained in the present
paper for the planar array evolve when one considers  the vertical
array of a {\em finite} thickness, $W$. Consider first the most
interesting case of intermediate thickness, $d \ll W \ll h$. The
modification of our result Eq.~(\ref{metalsol}) to this case can
be understood from the following reasoning. Each square with the
side $W$ contains $(W/d)^2$ NTs. This square can be viewed as a
single ``combined'' NT. Then the charge density in this combined
NT is given by Eq.~(\ref{metalsol}) with ${\cal N}_0$ replaced by
$W^{-1}$. To find  the charge density {\em per a single} NT, this
result should be divided by $(W/d)^2$.  Thus we conclude, that
modification of Eq.~(\ref{metalsol}) to finite $W<h$ amounts to
the replacement of ${\cal N}_0^{-1}$ by $d^2/W$, so that the
induced charge density falls off as $1/W$ with increasing the
width. However, the applicability of Eq.~(\ref{metalsol}) to the
finite-width array is limited by the size of a ``combined'' NT,
i.e., $(h-z)$ must exceed $W$. As $W$ exceeds the NT length, this
condition cannot be met, and the charge distribution changes
dramatically from the power-law behavior to the exponential
screening\cite{we2}.

Note that our basic integral equation, Eq.~(\ref{inteq1}),
expressing the fact that electrochemical potential is constant
along each NT of the array allow the exact solution using the
technique of conformal mapping only when the ``self-action''
[second term in the rhs of Eq.~(\ref{inteq1})] is neglected.  This
step is well justified for a dense array. In the electrostatics of
inhomogeneous fractional quantum Hall liquids
\cite{kogan92,kohmoto98} the charge distribution is described by
the equations similar to Eq.~(\ref{inteq1}), i.e., with
logarithmic kernel, but with self-action playing an important
role. Then these equations can be solved only numerically.


 \acknowledgements

This work was supported by   NSF under Grant No. DMR-0503172 and
by the Petroleum Research Fund under Grant No.  43966-AC10.

\vspace{4mm}

\section{Appendix}

Below we present a rigorous derivation of Eq.~(\ref{inteq1}) for a
periodic array of NTs. We start from the identity
\begin{eqnarray}
\label{2} \frac{1}{\sqrt{x^2+b^2}}=
\frac{1}{\pi}\int_{-\infty}^{\infty}\!\! dq\; K_0\Bigl(\vert
x\vert \;
q\Bigr)\exp\left(iqb \right),\nonumber\\
\end{eqnarray}
where $K_0(x)$ is the modified Bessel function of the second kind.
To use this identity in Eq.~(\ref{discrete}), we set $b\rightarrow
R_n=nd$. Then the summation in the integrand in the rhs can be
readily performed; it transforms the sum over $n$ in
Eq.~(\ref{kernel}) into the following sum of the
$\delta$-functions
\begin{equation}
\label{3} \sum_n \exp \left(iqR_n
\right)=\frac{2\pi}{d}\sum_{p}\delta\Biggl(q-\frac{2\pi
p}{d}\Biggr),
\end{equation}
where $p$ assumes {\em all} integer values. As the next step we
isolate the term with $p=0$ in the kernel Eq.~(\ref{kernel}) from
all other terms with nonzero $p$. Then we make use of the fact,
that
\begin{eqnarray}
\label{Macdonald} K_0(x)=-\gamma +\ln 2 -\ln\vert
x\vert+O(x^2),\qquad x\ll 1,
\end{eqnarray}
where $\gamma$ is the Euler constant. Then, for the term with
$p=0$,  we get
\begin{eqnarray}
\label{zero_p}{2\over d} \lim _{q\rightarrow 0}\bigl[K_0(\vert
z-z^{\prime}\vert q)-K_0(\vert z+z^{\prime}\vert q)\bigr]={2\over
d}\ln \Biggl|{\frac{ z-z^{\prime}}{
z+z^{\prime}}}\Biggr|.\nonumber\\
\end{eqnarray}
Therefore, after performing the summation over $n$ in
Eq.~(\ref{kernel}) and using Eq.~(\ref{zero_p}), the rhs of
Eq.~(\ref{discrete}) will acquire the form
\begin{equation}
\label{4} \int_0^{h}dz^{\prime}\rho(z^{\prime}){\cal
S}(z,z^{\prime}),
\end{equation}
where the kernel $S(z,z^{\prime})$ is given by
\begin{eqnarray}
\label{5} {\cal S}(z,z^{\prime})=\frac{2}{d}\ln \Biggl|{\frac{
z-z^{\prime}}{
z+z^{\prime}}}\Biggr|\qquad\qquad\qquad\qquad\qquad\\
+\frac{2}{d}\sum_{p\neq 0}\Biggl[K_0\Bigl(\vert
z-z^{\prime}\vert\; \frac{2 \pi p}{d}\Bigr)-K_0\Bigl(\vert
z+z^{\prime}\vert\; \frac{2 \pi p}{d}\Bigr)\Biggr].\nonumber
\end{eqnarray}
First term in (\ref{5}) describes the continuous limit; it
reproduces the second term in the rhs of Eq.~(\ref{inteq1})   and
comes from $p=0$ in Eq.~(\ref{3}).
The remaining sum over $p$ gives rise to the ``self-action'' term
in Eq.~(\ref{inteq1}). The easiest way to see this is to replace
the summation  over  $p$ by integration $dp$, which would
immediately yield
\begin{eqnarray}
\label{single} S_0(z,z^{\prime})=\phi (z-z^{\prime})-\phi
(z+z^{\prime})
\end{eqnarray}
for the second part of the kernel (\ref{5}). Note that
$S_0(z,z^{\prime})$ is nothing but the kernel for a {\em single}
NT~\cite{we1,we2}.
However, the replacement of the $\sum\limits _{p\neq 0}$ by $\int
dp$ is justified only when a large number of terms contribute to
the sum. This is the case only when the condition $\vert
z-z^{\prime}\vert \lesssim d$ is met. Then, upon substituting
(\ref{single}) into (\ref{3}) and restricting integration over
$z^{\prime}$ to $\vert z^{\prime}-z\vert \lesssim d$,   the
self-action term in Eq.~(\ref{inteq1}) is recovered. Concerning
the second part of the kernel~(\ref{5}) at $\vert
z^{\prime}-z\vert > d$, it is dominated by the terms with $p=\pm
1$, which behave as $\Bigl(d/\vert
z-z^{\prime}\vert\Bigr)^{1/2}\!\!\exp\;\Bigl\{-2\pi\vert
z-z^{\prime}\vert/d\Bigr\}$. Contribution coming from this part to
the integral Eq.~(\ref{4}) is $\sim \rho(z)$, which is
parametrically smaller than the self-action term $2{\cal L}_d
\rho(z)$.


\begin{references}





\bibitem{early0}  M. Terrones, N. Grobert, J. Olivares, J. P. Zhang, H. Terrones, K. Kordatos, W. K. Hsu, J. P. Hare,
P. D. Townsend, K. Prassides, A. K. Cheetham, H. W. Kroto, and D.
R. M. Walton, Nature (London) {\bf 388}, 52 (1997).

\bibitem{early2} Z. F. Ren, Z. P. Huang, J. W. Xu, J. H. Wang, P. Bush, M. P. Siegal,
and P. N. Provencio, Science {\bf 282}, 1105 (1998).



\bibitem{early1}Q. H. Wang, A. A. Setlur, J. M. Lauerhaas, J. Y. Dai, E. W. Seelig, and R. P. H. Chang,
Appl. Phys. Lett. {\bf 72}, 2912 (1998).


\bibitem{early3} S. Fan, M. G. Chapline, N. R. Franklin, T. W. Tombler, A. M. Cassell,
and H. Dai, Science {\bf 283}, 512 (1999).


\bibitem{early4}O. Gr\"{o}ning, O. M. K\"{u}ttel, Ch. Emmenegger, P. Gr\"{o}ning, and L. Schlapbach,
 J. Vac. Sci. Technol. B {\bf 18}, 665 (2000).

\bibitem{early5} L. Nilsson, O. Groening, C. Emmenegger, O. Kuettel, E. Schaller, L. Schlapbach, H. Kind, J.-M. Bonard,
and K. Kern, Appl. Phys. Lett. {\bf 76}, 2071 (2000).

\bibitem{early6} J.-M. Bonard, K. A. Dean, B. F. Coll, and C. Klinke,
Phys. Rev. Lett. {\bf 89}, 197602 (2002).

\bibitem{sensing} F. Picaud, R. Langlet, M. Arab, M. Devel, C. Girardet,
S. Natarajan, S. Chopra, and A. M. Rao, J. Appl. Phys. {\bf 97},
114316 (2005).





\bibitem{latest1}  V. I. Merkulov, D. H. Lowndes, and L. R. Baylor,
J. Appl. Phys. {\bf 89}, 1933 (2001).

\bibitem{latest3}
M. Chhowalla, C. Ducati, N. L. Rupesinghe, K. B. K. Teo, and G. A.
J. Amaratunga, Appl. Phys. Lett. {\bf 79}, 2079 (2001).

\bibitem{latest2} J. S. Suh, K. S. Jeong,  J. S. Lee,
and I. Han,    Appl. Phys. Lett. {\bf 80}, 2392 (2002).

\bibitem{latest4}
H. J. Lee, S. I. Moon,  J. K. Kim, Y. D. Lee,  S. Nahm, J. E. Yoo,
J. H. Han, Y. H. Lee, S. W. Hwang, and B. K. Ju, J.~Appl. Phys.
{\bf 98}, 016107 (2005).





\bibitem{growth} A. Ural, Y. Li, and H. Dai ,
Appl. Phys. Lett. {\bf 81}, 3464 (2002).

\bibitem{growth1} A. Nojeh, A. Ural, R. F. Pease, and H. Dai, J.
Vac. Sci. Technol. B {\bf 22}, 3421 (2004).


\bibitem{ajayan00} Z. J. Zhang, B. Q. Wei, G. Ramanath, and P. M.
Ajayan, Appl. Phys. Lett. {\bf 77}, 3764 (2000).

\bibitem{ajayan01} Z. J. Zhang, P. M. Ajayan, G. Ramanath,  J. Vacik and Y. H. Xu,
Appl. Phys. Lett. {\bf 78}, 3794 (2001).

\bibitem{ajayan02} A. Cao, B. Wei, Y. Jung, R. Vajtai, P. M. Ajayan, and G. Ramanath,
 Appl. Phys. Lett. {\bf 81}, 1297 (2002).

\bibitem{iowa04}V. V. Tsukruk, H. Ko, and S. Peleshanko,  Phys. Rev. Lett. {\bf 92}, 065502 (2004).



\bibitem{Nilsson}
L. Nilsson, O. Groening, C. Emmenegger, O. Kuettel, E. Schaller,
L. Schlapbach, H. Kind, J.-M. Bonard, and K. Kern, Appl. Phys.
Lett. {\bf 76}, 2071 (2000).

\bibitem{we2} T. A. Sedrakyan, E. G. Mishchenko, and M. E. Raikh,
Phys. Rev. B {\bf 73}, 245325 (2006).

\bibitem{wang05}
M. Wang, Z. H. Li, X. F. Shang, X. Q. Wang, and Y. B. Xu,
 J. Appl. Phys. {\bf 98}, 014315 (2005).




\bibitem{we1} E.~G.~Mishchenko and M.~E.~Raikh,
Phys. Rev. B {\bf 74}, 155410 (2006).


\bibitem{ThomasFermi} M. P. Anantram and  F. L\'{e}onard,
 Rep. Prog. Phys. {\bf 69}, 507 (2006).




\bibitem{muskhelishvili} N. I. Muskhelishvili,
{\em Singular Integral Equations; Boundary Problems of Function
Theory and Their Application to Mathematical Physics}, Groningen,
P. Noordhoff, 1953.

\bibitem{nehari} Z. Nehari, {\em  Conformal mapping
 }, New York, McGraw-Hill, 1952.

\bibitem{schwarz-chrisoffel} T. A. Driscoll and L. N. Trefethen,
{\em  Schwarz-Christoffel Mapping}, Cambridge University Press,
2002.

\bibitem{conf-map} V. I. Ivanov and M. K. Trubetskov,
{\em Handbook of Conformal Mapping with Computer-Aided
Visualization}, CRC Press, 1995.

\bibitem{glazman91} L. I. Glazman and I. A. Larkin, Semicond. Sci. Technol. {\bf 6},
32 (1991).

\bibitem{chklovskii} D. B. Chklovskii, B. I. Shklovskii, and L. I. Glazman,
Phys. Rev. B {\bf 46}, 4026 (1992).



\bibitem{kennedy05} J. W. Kennedy, Z. V. Vardeny, S. Collins, R. H. Baughman, H. Zhao,
S. Mazumdar, preprint cond-mat/0505071.




\bibitem{Kane}C. L. Kane and E. J. Mele,
 Phys. Rev. Lett. {\bf 90}, 207401 (2003);
{\bf 93}, 197402 (2004).
\bibitem{Spataru} C. D. Spataru, S. Ismail-Beigi, L. X. Benedict, and S. G. Louie,
Phys. Rev. Lett. {\bf 92}, 077402 (2004);
\bibitem{Perebeinos} V. Perebeinos, J. Tersoff, and P. Avouris, Phys. Rev. Lett.
{\bf 92}, 257402 (2004).


\bibitem{Liang} W.-Z. Liang,
G.-H. Chen, Z. Li, and Z.-K. Tang, Appl. Phys. Lett. {\bf 80},
3415 (2002).

\bibitem{exciton}H. Zhao and S. Mazumdar,
 Phys. Rev. Lett. {\bf 93}, 157402 (2004).




\bibitem{Landau84} L. D. Landau and E. M. Lifshitz, {\em Electrodynamics of Continuous Media}
(Pergamon Press, Oxford, 1984).





\bibitem{network1} E. S. Snow, J. P. Novak, P. M. Campbell, and D.
Park, Appl. Phys. Lett. {\bf 82}, 2145 (2003).

\bibitem{network2} E. S. Snow, P. M. Campbell, M. G. Ancona, and J. P.
Novak, Appl. Phys. Lett. {\bf 86}, 033105 (2005).

\bibitem{network3} S. Kumar, N. Pimparkar, J. Y. Murthy, and M. A.
Alam, Appl. Phys. Lett. {\bf 88}, 123505 (2006).






\bibitem{kogan92} I. Kogan, A. M. Perelomov, and G. W. Semenoff,
Phys. Rev. B {\bf 45}, 12084 (1992).




\bibitem{kohmoto98} J. Shiraishi, Y. Avishai, and M. Kohmoto,
Phys. Rev. B {\bf 57}, 13061 (1998).




\end{references}
\end{document}